\documentclass[%
 reprint,
 amsmath,amssymb,
 aps,
]{revtex4-1}

\usepackage{url}
\usepackage{float}
\usepackage{graphicx}% Include figure files
\usepackage{dcolumn}% Align table columns on decimal point
\usepackage{bm}% bold math
\begin{document}

%\preprint{APS/123-QED}

\title{Simulation of the Kinetic Energy Spectrum of Primary Electron \\ Generated by a 6 keV X-Ray Source in a Gas Volume}% Force line breaks with \\
%% ============== Gamma Beam --> X-Ray Source
%% ============== Gaseous Medium --> Gas Volume

\author{Lucas Maia Rios}
\email{lucas.maia@aluno.ufabc.edu.br}

\author{Mauro Rogerio Cosentino}
\email{mauro.cosentino@ufabc.edu.br}
\affiliation{%
 Federal University of ABC
}%

\date{May 18, 2018}% It is always \today, today,

\begin{abstract}

With the intention of creating a complete simulation of the Thick GEM detector, this work presents a partial simulation of the interaction of 6 keV X-Ray radiation with a gas volume. For the gas, it was used pure Argon, Xenon and CO2 as well as a mixture of each noble gas with CO2 in a 70\%-30\% proportion. For this range of energy, the dominant physical effect is the photoelectric, therefore knowing the binding energies of the atoms we predict the kinetic energy of the primary electron. A simulated energy spectrum of the primary electron kinetic energy was obtained, which contains peaks corresponding to the inner shell's ionization of the atoms. The spectrum shows very little Compton effect, transferring only small energies (in the order of eVs). We concluded that energy spectrum for the mixture is a linear combination of the spectrum of its individual components.

. 
\begin{description}
\item[Keywords]
electron energy spectrum, X-Ray beam, simulation, Geant4.
\end{description}
\end{abstract}

\pacs{Valid PACS appear here}% PACS, the Physics and Astronomy
\maketitle

%\tableofcontents

\section{Introduction}

\subsection{Motivation}

Gaseous detectors are present in the fields of nuclear and particle physics since the beginning, but it was only after the invention of the Multi-Wire Proportional Chamber (MWPC) that they became present in almost every experiment. There are applications for them in medical science, engineering and even in every day use technology. The application we are interested in to further study is the detection of charged particles created in a particle accelerator collision. The ALICE experiment (A Large Ion Collider Experiment) is a multi-detector experiment at the LHC at CERN that is designed to detect particles generated in heavy-ion collisions. The goal of the experiment is to study the quark-gluon plasma, which is studied indirectly through the decay modes of the plasma.

To detect these particles, it is currently in use the MWPC technology, within a Time Projection Chamber (TPC) \cite{tpc_alice}, aimed to reconstruct the track of charged particles. A MWPC is a chamber with anode wires between cathode planes arranged in such a way to create an electric field that captures the electrons from ionizing radiation. The electron travels through the drift area, to the multiplication area where the electric field intensifies and triggers an avalanche of electrons that induce a signal in the wires.

One problem of using MWPC though, is that it creates a significant ion back flow (IBF) as a result of the multiplication. As the ions are heavier than the electrons, they are much slower. Therefore, they accumulate in the drift area. This concentration of ions distorts the electric field, thus distorting the tracking reconstruction. To prevent this a gating grid is set in place, in order to block the passage of the ions, but it also prevents electrons from the drift area to reach the anode wires. Therefore, the gating grids must be opened for a period at each event, short enough to prevent the IBF but long enough to allow the electrons to go through.

In 2019 the LHC will undergo a shutdown for upgrades on the luminosity of the beam (to increase the number of collisions per second) as well as other improvements in the experiments. As for the ALICE experiment upgrade \cite{alice_upgrade}, it is expected to tackle this IBF issue by switching the MWPC to GEM (Gas Electron Multiplier) \cite{Sauli} in the TPC, since the higher collision rate makes it impossible for the gating grid to work correctly.

\begin{figure}[h!]
\centering
\includegraphics[scale=0.4]{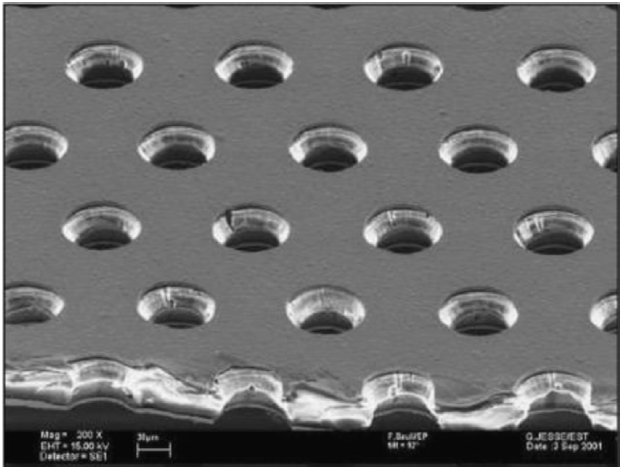}
\caption{Typical GEM electrode, 50 $\mu m$ thick, 140 $\mu m$ pitch and  70 $\mu m$ diameter. \cite{Sauli}}
\label{fig:GEM_foil}
\end{figure}

\begin{figure}[h!]
\centering
\includegraphics[scale=0.4]{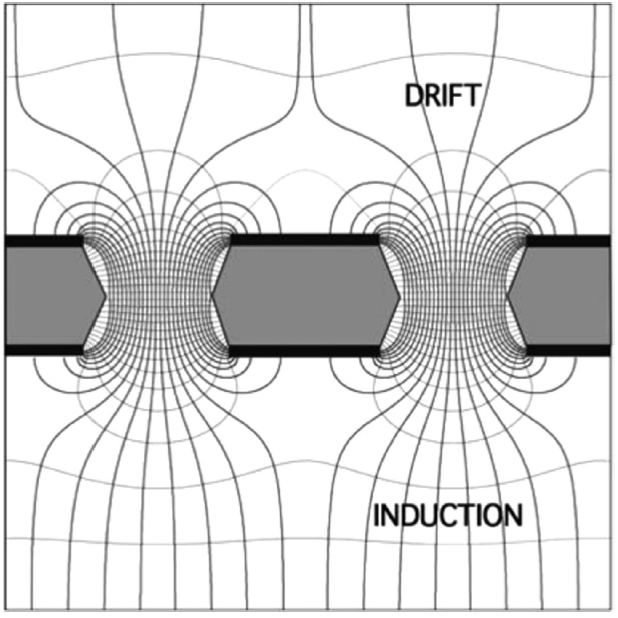}
\caption{Electric field in the region of the holes. \cite{Sauli}}
\label{fig:GEM_ef}
\end{figure}

GEMs are a pair of foils with a dieletric layer in between, with equally spaced holes as shown in figure \ref{fig:GEM_foil}. There is an electric potential difference between these two foils so it creates an intense electric field in the holes. The charged particles will tend to follow the eletric field lines as shown in figure \ref{fig:GEM_ef}. The primary electron from an ionization travels through the equipotential lines in the drift area until it reaches the multiplication area inside the hole. On the other side of the GEM comes out a greater number of electrons. Whereas ions follow the opposite path through the equipotential lines to either return to the drift area or to be deposited in the GEM foil, reducing the IBF. Also, one can arrange several GEMs in series so the ions created are captured by the previous GEM. Only a small fraction of the ions go to the drift area.

GEMs are considerably expensive and sensitive (a short-circuit could damage them). For that reason, we have been developing a Thick GEM prototype. Conceptually there is no difference between the two. However, Thick GEMs are more robust. Their holes are significantly larger and the dieletric layer is thicker (this is why they are called Thick GEM). The disadvantage is the reduced position resolution.

Thick GEMs, for their low price and better efficiency than MWPC, could be used for several applications. Here at Federal University of ABC, there are laboratories that work with X-Ray crystallography that still use MWPC that could benefit from the use of the Thick GEM.

This work is an ongoing project to develop a complete simulation of an incident monochromatic X-Ray beam, ionization the gas and then detection of the electrons by the Thick GEM. In order to have the complete simulation, three softwares will be needed. Geant4 \cite{geant4} is used to simulate the interaction of the radiation with matter. Then Garfield++ \cite{garfieldpp} will be used to simulate the electron transport, multiplication and signal generation. At last, ROOT \cite{root} is used to analyze the data.

This paper contains the simulation of the radiation interaction with the gas and a study of the electron spectrum created by it.

\subsection{Physics}

Ionizing radiation can interact with matter through many ways. Some of which result in the creation of charged particles. The particle of interest in this study that will be interacting is the photon. A source of X-ray radiation emits photons that, when compared to charged particles, can better penetrate a material because of its neutral nature. Nevertheless, it may interact with the electrons of the material along its path. The physical processes that a photon can undergo are the photoelectric effect, Compton scattering, pair production, Thomson and Rayleigh scattering. The latter two are non-ionizing elastic scattering, thus they will not be discussed. Also, for the pair production to occur it is necessary that the photon has an energy greater than a certain threshold (at least the rest mass energy of the produced particles). The rest energy of an electron and a positron together is 1022 keV. At the energy we are working there will be no pair production, so we will put it aside as well. That leaves us with photoelectric effect and Compton scattering, both transfer energy to the medium that manifests as an electron gaining kinetic energy. In the first case the photon is completely absorbed by an atom, which ejects an electron as a result. In the latter the photon only transfers a fraction of its energy, allowing it to further interact via either one of the processes or to escape the volume of interest, without further interactions.

The electron bounded to an atom is trapped within a potential barrier. This barrier is interpreted as the binding energy. In an interaction with ionizing radiation it is more likely that the innermost electron in the atom is freed. \cite{leo} With this in mind, we can write the kinetic energy of an electron ejected via photoelectric effect as:

\begin{equation}
    K_e = E_{\gamma} - \phi,
\label{eq:photo}
\end{equation}

where $ E_{\gamma}$ is the incoming photon energy and $\phi$ the binding energy.

Given the atoms in the gaseous medium, and knowing its binding energies, we can foretell the kinetic energy of the ejected electron. Also note that it will be a discrete spectrum, since the binding energies are discrete and our X-ray ray is monochromatic. The binding energy for the innermost electron in Ar-40, for example, is 3205.9 eV \cite{xray2}. Therefore, using equation \ref{eq:photo} we expect our electron to have a kinetic energy of 2794.1 eV. All of the binding energies for Argon and Xenon can be seen in table \ref{tab:be}, as well as the expected kinetic energy for the electron.

\begin{table}[]
\centering
\label{b_table}
\begin{tabular}{|c|c|c|c|}
\hline
                  & Ar     & Xe     & Kinetic E. ($eV$) \\ \hline
$K$ 1s            & 3205.9 & 34561  &  2794.1 /    -       \\ \hline
$L_1$ 2s          & 326.3  & 5453   &  5673.7 / 547      \\ \hline
$L_2$ 2p$_{1/2}$  & 250.6  & 5107   &  5749.4 / 893        \\ \hline
$L_3$ 2p$_{3/2}$  & 248.4  & 4786   &  5751.6 / 1214           \\ \hline
$M_1$ 3s          & 29.3   & 1148.7 &  5970.7 / 4851.3           \\ \hline
$M_2$ 3p$_{1/2}$  & 15.9   & 1002.2 &  5984.1 / 4997.8           \\ \hline
$M_3$ 3p$_{3/2}$  & 15.7   & 940.6  &  5984.3 / 5059.4          \\ \hline
$M_4$ 3d$_{3/2}$  &        & 689.0  &   5311.0  \\ \hline
$M_5$ 3d$_{5/2}$  &        & 676.4  &   5323.6          \\ \hline
$N_1$ 4s          &        & 213.2  &   5786.8          \\ \hline
$N_2$ 4p$_{1/2}$  &        & 146.7  &   5853.3          \\ \hline
$N_3$ 4p$_{3/2}$  &        & 145.5  &   5854.5          \\ \hline
$N_4$ 4d$_{3/2}$  &        & 69.5   &   5930.5          \\ \hline
$N_5$ 4d$_{5/2}$  &        & 67.5   &   5932.5          \\ \hline
$O_1$ 5s          &        & 23.3   &   5976.7          \\ \hline
$O_2$ 5p$_{1/2}$  &        & 13.4   &   5986.6          \\ \hline
$O_3$ 5p$_{3/2}$  &        & 12.1   &   5987.9          \\ \hline
\end{tabular}
\caption{Binding energy for Ar and Xe and their respective electron kinetic energy when ejected by 6 $keV$ photon. Data collected from X-Ray Data Booklet which used \cite{xray1} and \cite{xray2}.}
\label{tab:be}
\end{table}

The Compton scattering, on the other hand, is a collision between the photon and a free electron. The amount of energy transferred is random, we can not know it exactly beforehand. However, we know that the loss of energy by the photon is completely gained by the electron as kinetic energy. 

Differently from the photoelectric effect, the electrons will gain kinetic energy within a continuous range of energy. Also, free electrons in gases are not as abundant as in metal. According to \cite{pdg} the Compton cross section for this range of energy is smaller than the photoelectric by 2 orders of magnitude. So it is expect that the latter should be the dominant process for 6 keV photons.

Our interest in this work is to simulate the X-ray interaction with a gas volume. We chose an energy of 6 keV because of the common use of Fe-55 as a source of X-Ray radiation, which has energy emissions about 6 keV. The reason for simulating X-Ray is due to the prospect of implementing the Thick GEM in the X-Ray crystallography laboratory. We expect to get as an output the electron energy spectra for different gas mixtures. From that we can compare with the theoretical predictions to test the reliability of our simulation.

\section{Simulation}

We used the Geant4 toolkit \cite{geant4}. It is an object-oriented C++ library of classes for simulation of radiation interaction with matter. With it we can define the geometry of our detector and its materials. For the primary event, the first particle created that is, the photon is defined. Its momentum direction is centered in the origin of the XY surface and it has a normal distribution along the Z axis of standard deviation of $\sigma = 1/12 \pi$ rad.

Considering the discussion presented in the previous section, the processes added to the simulation were the photoelectric effect and the Compton scattering.

The initial purpose of Geant4 was simulating high energy physics. Because of that there has been for some time a deficit in quality for low energy interaction. In recent years, however, new models have been implemented such as Livermore and Penelope. Both have a low limit applicability of 250 eV. For the high limits they are 100 GeV and 1 GeV, respectively. %We considered both models for comparison and verify which one better fits the theoretical prediction, if there's any difference between the models.%

The primary event, i.e. the first particle (photons from our X-Ray source) of the simulation, is generated in the origin, immerse in air. In 1 mm of distance of the origin in the Z axis there is a volume filled with a gas mixture (Argon-CO2 or Xenon-CO2, in 70\%-30\% or its pure constituents). This volume is a cube of 5 cm of side.  All of the physics of interest is in this volume. The X-ray ray will interact with the particles of such gas and deposit energy in it. It will also ionize it, creating electrons and ions.

In the simulation a number of events is set (each event is the creation of the primary particle until it and all its secondaries are absorbed and there no longer are particles to track). That number is currently ten thousand. A particle is tracked step-by-step, in which we know all of its information (position, momentum, energy, etc) before and after the current step. Every time a particle penetrates the volume of our interest, information about the hits (interactions) is collected. If there are electrons created by the X-ray, their kinetic energy is saved as well.

\begin{figure}[h!]
\centering
\includegraphics[scale=0.41]{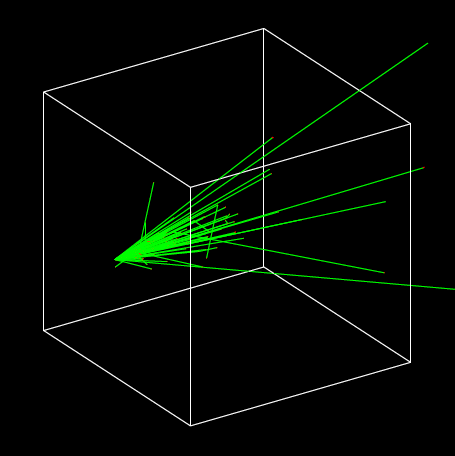}
\caption{Perspective view for Argon 70\%. Photon trajectory in green.}
\label{fig:p_view}
\end{figure}

\begin{figure}[h!]
\centering
\includegraphics[scale=0.4]{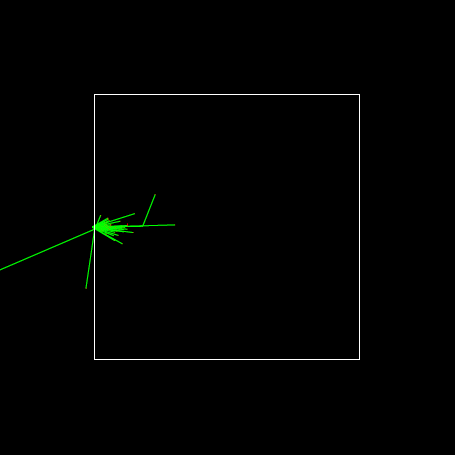}
\caption{Side view for Xenon 70\%. Photon trajectory in green.}
\label{fig:side_view}
\end{figure}

Geant4 also has visualization tools as we can see in figure \ref{fig:p_view} and \ref{fig:side_view}. They both show the primary particle traveling into the volume and interacting with its gas. In green is the photon. The primary electrons were to be seen in red, however, because of their low energy they are quickly absorbed by the medium and barely travel.

In the side view (figure \ref{fig:side_view}), we can visualize the collimation of the beam. Another difference in the two figures is the composition of the gas. We can see that because Xenon has a higher Z number the X-Ray travels shorter distances than in Argon.

\section{Results}

The output of the Geant4 simulation is a ROOT data file which contains the information of position, momentum of the interactions as well as the energy spectrum of the primary electron. We ran the simulation for the pure noble gases (Ar and Xe), CO2 and their mixtures with 70\% of noble gas and 30\% of CO2. In a histogram there are 200 channels per keV. We begin by looking at spectrum created by Argon alone.

\begin{figure}[h!]
\centering
\includegraphics[scale=0.4]{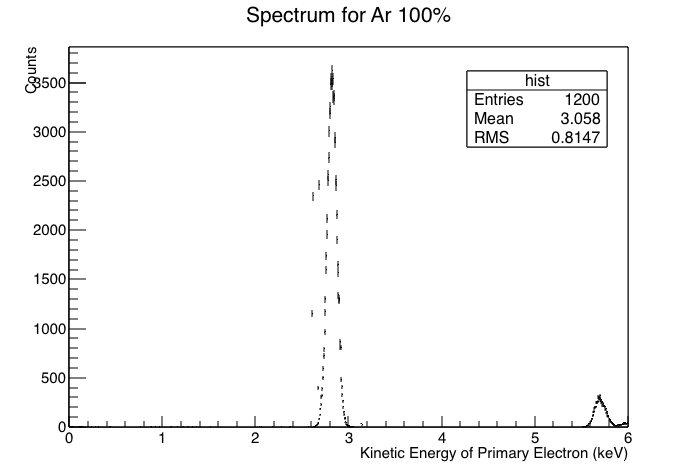}
\caption{Primary electron energy spectrum for Argon 100\%.}
\label{fig:Ar100_01}
\end{figure}

\begin{figure}[h!]
\centering
\includegraphics[scale=0.4]{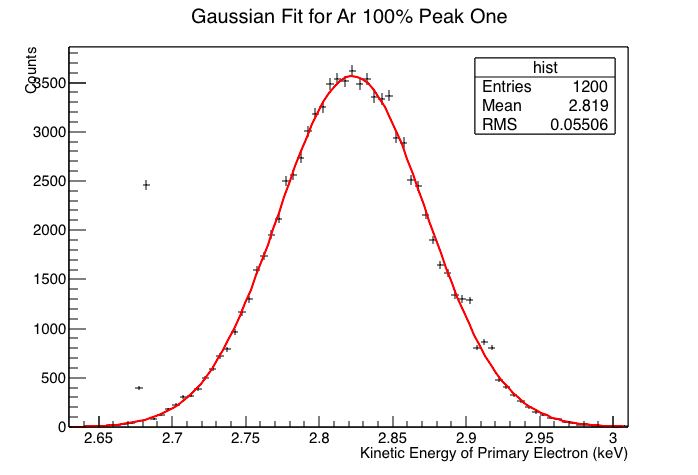}
\caption{Peak one for Argon 100\%.}
\label{fig:Ar100_Peak01}
\end{figure}

\begin{figure}[h!]
\centering
\includegraphics[scale=0.4]{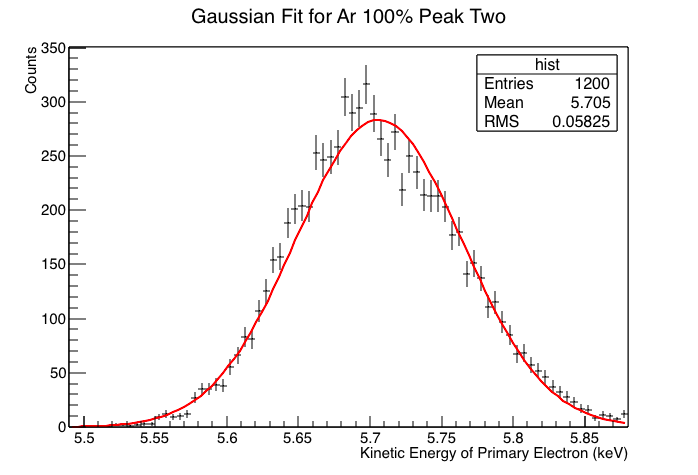}
\caption{Peak two for Argon 100\%.}
\label{fig:Ar100_Peak02}
\end{figure}

In figure \ref{fig:Ar100_01} we can see that there are two visible peaks. One is in the middle of the range and the other one is closer to 6 keV. If we take a closer look at the peaks, figures \ref{fig:Ar100_Peak01} and \ref{fig:Ar100_Peak02}, we can fit a gaussian in each peak using ROOT. As a result their mean values and errors are 2.8228 $\pm$ 0.0002 keV and 5.7061 $\pm$ 0.0007 keV. For each peak from here on the fitting method will be used to get their information.

We expected that the main energy peaks would correspond to inner shells of the atoms. From table \ref{tab:be} the energies for the K and L shell for Argon are rather close to the peaks. That is, K shell has a 1.02\% relative difference. For the L shell however, it seems that there could be an overlap of the L shell peaks in the spectrum because we see only one peak instead of three. Nonetheless, the relative differences are respectively 0.57, 0.75 and 0.79\%.

\begin{figure}[h!]
\centering
\includegraphics[scale=0.4]{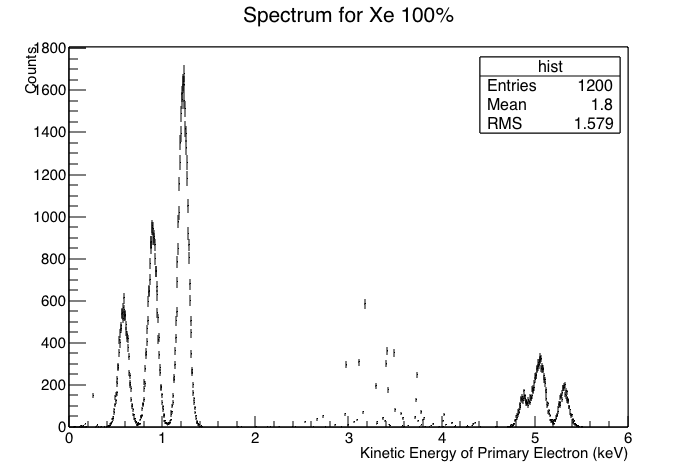}
\caption{Primary electron energy spectrum for Xenon 100\%.}
\label{fig:Xe100_01}
\end{figure}

As for the Xenon gas alone, we can see two sets of three peaks (figure \ref{fig:Xe100_01}). The first set is between 0 and 2 keV and their mean values are 0.5830 $\pm$ 0.0004, 0.8960 $\pm$ 0.0003 and 1.2252 $\pm$ 0.0002 keV. For the second set there is some overlapping but we can still see the three different peaks with mean values of 4.881 $\pm$ 0.002, 5.052 $\pm$ 0.001 and 5.3160 $\pm$ 0.0008 keV.

\begin{figure}[h!]
\centering
\includegraphics[scale=0.4]{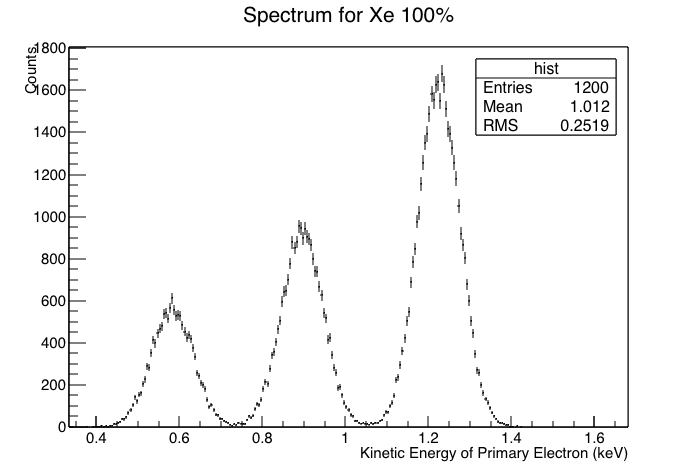}
\caption{Lower energy peaks for Xenon 100\%.}
\label{fig:Xe100_02}
\end{figure}

\begin{figure}[h!]
\centering
\includegraphics[scale=0.4]{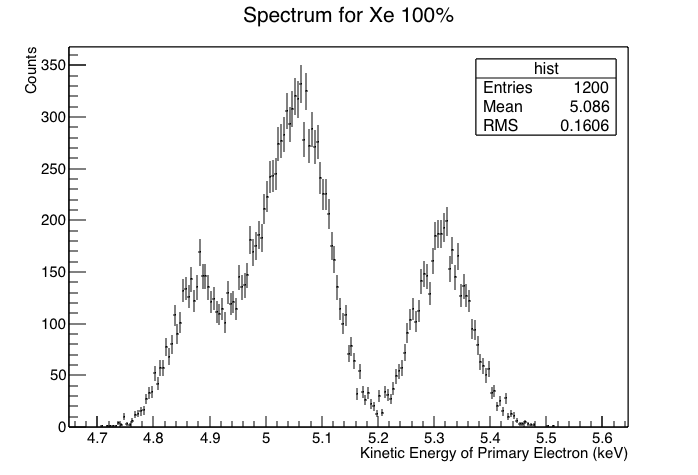}
\caption{Higher energy peaks for Xenon 100\%.}
\label{fig:Xe100_03}
\end{figure}

From table \ref{tab:be} we can infer that the first three peaks correspond to the three L shell energies. Since the photon does not have enough energy to ionize the K shell, L shell becomes the innermost ionizable one. The relative difference to the peaks are respectively 6.58, 0.33 and 0.92\%. The second set of peaks correspond to the next innermost shell, the M shell. Their relative differences are 0.61 ($M_1$), 0.15 ($M_3$) and 0.09 \% ($M_4$) to the theoretical prediction.

\begin{figure}[h!]
\centering
\includegraphics[scale=0.4]{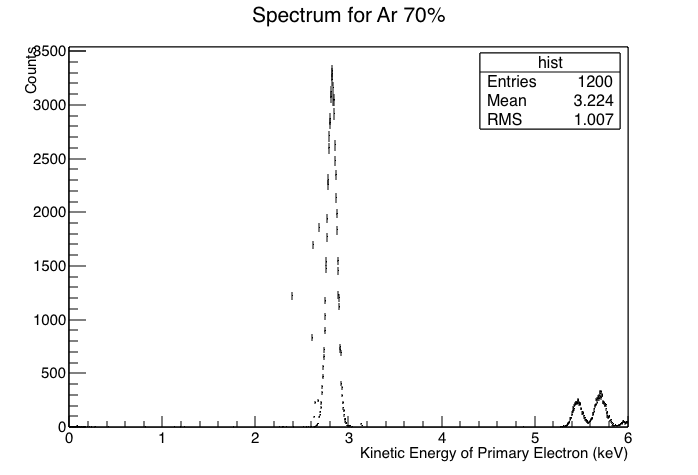}
\caption{Primary electron energy spectrum for Argon 70\%.}
\label{fig:Ar70_01}
\end{figure}

\begin{figure}[h!]
\centering
\includegraphics[scale=0.4]{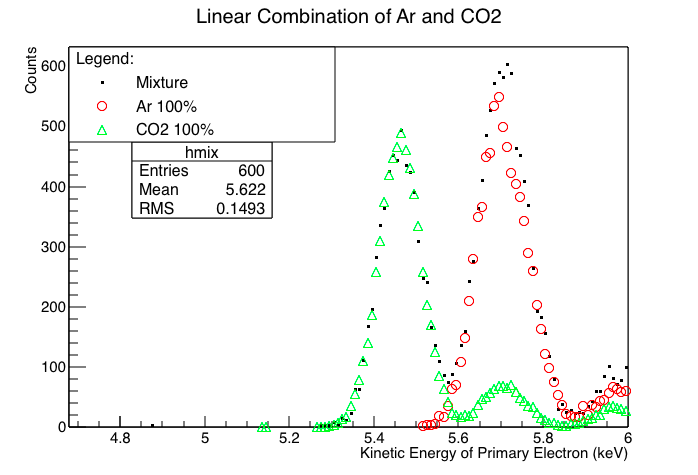}
\caption{In black there is the simulated spectrum for the mixture 70\% Argon and 30\% CO2. In red the pure Argon spectrum rescaled by 0.9 whereas in green is the pure CO2 rescaled by 0.2.}
\label{fig:Ar_mix}
\end{figure}

Now we can start looking at the mixtures.  First for Argon 70\% with CO2 30\%, in figure \ref{fig:Ar70_01}. We expect that the same peaks will appear, changing only their amplitudes. The histogram of mix thus should be a linear combination of their pure histograms. From figure \ref{fig:Ar70_01} we can see that the peak for the K shell remains. However, there is a distortion where the L shell would be due to the presence of CO2, as we can see in figure \ref{fig:Ar_mix}. We have plotted in figure \ref{fig:Ar_mix} both the mixture and its individual components rescaled to show that when summed they become the mixture histogram.

\begin{figure}[h!]
\centering
\includegraphics[scale=0.4]{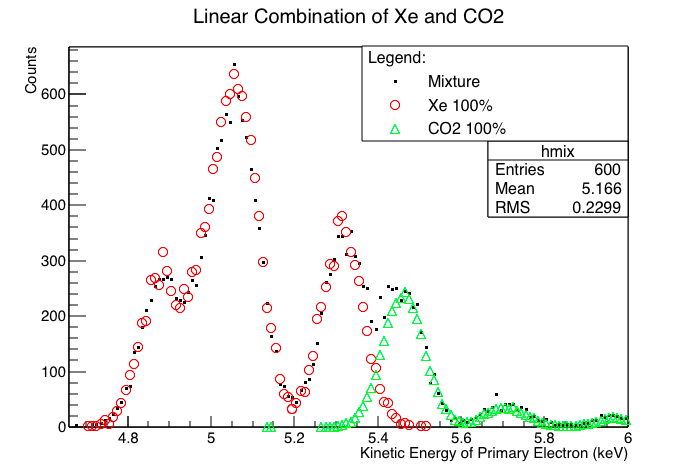}
\caption{In black there is the simulated spectrum for the mixture 70\% Xenon and 30\% CO2. In red the pure Xenon spectrum whereas in green is the pure CO2 spectrum rescaled by 0.1.}
\label{fig:Xe_mix}
\end{figure}

Identically we can do the same for the Xenon 70\% mixture in figure \ref{fig:Xe_mix}.

\begin{figure}[h!]
\centering
\includegraphics[scale=0.4]{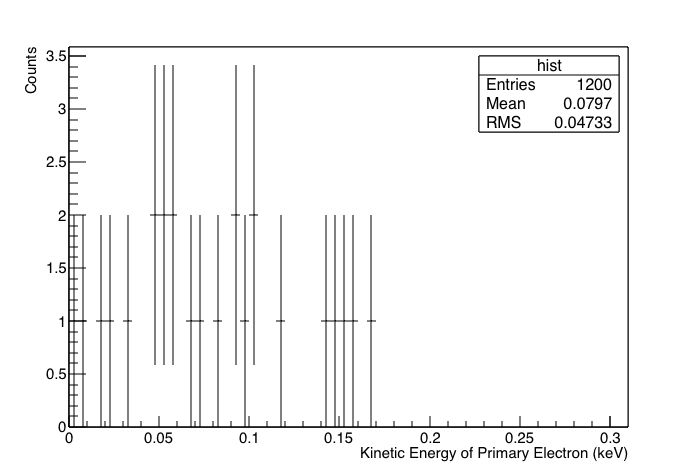}
\caption{Compton scattering spectrum for Xenon 70\%.}
\label{fig:Xe_Compt}
\end{figure}

For Argon, CO2 and their mixture there are nothing else to explore in the spectrum. For the Xenon, however, there are some counts that do not belong to the Xenon binding energies because there are no shells between the L and M that correspond to that energy (3-4 keV) as we can see in figure \ref{fig:Xe100_01}. One could think those energy counts are due to Compton scattering, but figure \ref{fig:Xe_Compt} show us that it only happens for energy in the order of eVs.

So according to the simulation those middle range counts should be Photoeletric effect. It could be some mistake in the simulation. More investigation needs to be done.

It should also be acknowledged that the measurement error depends of the standard deviation we choose for the energy. Since the Fe-55 has different energy emissions, we approximated the energy to a Gaussian with mean energy of 6 keV.

\section{Conclusion}

The simulation is able to create the energy peaks corresponding to the binding energies of the gas filling the volume. The highest relative difference being 6.58\%, however most of the rest are less than 1\%. As the error associated to the measurement in the simulation is mostly in the order of $10^{-2}$ or $10^{-3}$\%, the relative difference can not be ignored. Also, we can conclude that the spectrum of the mixtures of gases is a linear combination of the spectrum of its individual components.

Further study of the simulation is necessary to grasp the reason why there are some scattered energy counts where there should not be. Experiments will be made in the future with the Thick GEM, then it will be possible to confront experimental data with the simulation.

%[] http://garfield.web.cern.ch/

\bibliography{main}{}
\bibliographystyle{plain}

\end{document}